\documentclass[prl,aps,twocolumn,superscriptaddress,amsmath,amssymb]{revtex4}
\usepackage{amsmath}
\usepackage[english]{babel}
\usepackage{graphicx}
\usepackage{bm}

\begin{document}

%
 \title{Giant Mesoscopic Fluctuations and Long Range Superconducting Correlations
 in Superconductor--Ferromagnet structures}

\author{A.~S.~Mel'nikov}
\affiliation{Institute for Physics of Microstructures, Russian
Academy of Sciences, 603950 Nizhny Novgorod, GSP-105, Russia}
\affiliation{
Lobachevsky State University of Nizhny Novgorod, 23 Gagarina, 603950 Nizhny Novgorod, Russia}
\author{A.~I.~Buzdin} \affiliation{University Bordeaux, LOMA UMR-CNRS 5798, F-33405
Talence Cedex, France}

\begin{abstract}
The fluctuating superconducting correlations emerging in dirty
hybrid structures in conditions of the strong proximity effect
are demonstrated to affect the validity range of the widely used formalism of Usadel equations at mesoscopic scales.
In superconductor -- ferromagnet (SF) structures these giant mesoscopic fluctuations originating from the interference effects for the Cooper pair wave function
in the presence of the exchange field can be responsible for an anomalously slow decay of superconducting correlations in a ferromagnet even when the
non-collinear and spin-orbit effects are negligible.
The resulting sample-to-sample fluctuations of the Josephson current in SFS junctions and local density of states in SF hybrid structures can
 provide an explanation of the long range proximity phenomena observed in mesoscopic samples with collinear magnetization.
\end{abstract}

\maketitle

The successful development of modern experiment in a wide class of superconductor (S) -- ferromagnet (F)
hybrid structures has opened a completely new research area, i.e., superconducting spintronics \cite{Linder,Eschrig,Eschrig2,pi}.
Despite the growing number of various exciting experimental and theoretical results in this field
there remains still a very important and puzzling contradiction between the experimental data and the understanding of the
physics of the proximity effect in these systems. This contradiction relates to the length of decay of superconducting correlations,
$L_s$,
in a dirty ferromagnet. The standard Usadel theory gives us the exponential suppression of superconducting correlations at the length
 $L_s\sim \xi_f=\sqrt{\hbar D/2h}$, where $D$ is the diffusion coefficient and $h$ is the exchange field (see \cite{buzdin} for review). For most typical ferromagnets and alloys
 one can get the estimate $L_s\sim 1-10$~nm while the existing experimental data \cite{giroud,petrashov,nugent,sosnin,keiser,rob,khaire,wang,dobr1,dobr2} show that superconducting correlations survive for much larger
 distances from the SF interface comparable with the coherence length in the normal metal $\xi_N\sim\sqrt{D/T}$. This puzzling discrepancy between
 the theoretical estimate and experiment stimulated researchers to suggest various fascinating mechanisms of such long range proximity phenomenon.
  Most of these explanations exploit the assumption of the inhomogeneous exchange field which generates the long range triplet
 pairing component in the anomalous Green functions (see, e.g., \cite{bergeret}). The resulting increase in the $L_s$ length has been confirmed experimentally for Josephson junction with a composite F layer containing a region with  non-collinear magnetic  moments \cite{rob,khaire}. Thus, the non-collinearity of magnetic structure nicely explains the overwhelming majority of the above experiments. An alternative source of the long range effect originating from the spin-orbit interaction has been recently discussed in Refs.~\cite{mel,bergeret1,bergeret2,linder}.
  However, up to now there are no satisfactory explanations of the experiments \cite{giroud,petrashov,nugent,wang}
where no traces of a non-collinear magnetization were reported and the strength of the spin-orbit effects can be a subject of debates.

Motivated by the above discrepancy between the experiment and theory we suggest to reexamine the
standard Usadel-type model and search for possible shortcomings of this model which can reveal themselves in the $L_s$ estimates.
One of the most important assumptions which form the basis of the Usadel theory is that we
operate with the ensemble-averaged Green functions neglecting, thus, possible fluctuations of the measurable quantities due to the random
distribution of impurities \cite{altland,skvortsov,zyuzin}. In the case of the dirty ferromagnet this assumption is crucial to get the exponential decay of the anomalous Green function
at the length $\xi_f$. Indeed, the motion of quasiparticles in a ferromagnetic metal occurs along the random quasiclassical trajectories
which experience sharp turns at the impurity positions (see Fig.~1).
\begin{figure}[t!]
\includegraphics[width=0.25\textwidth]{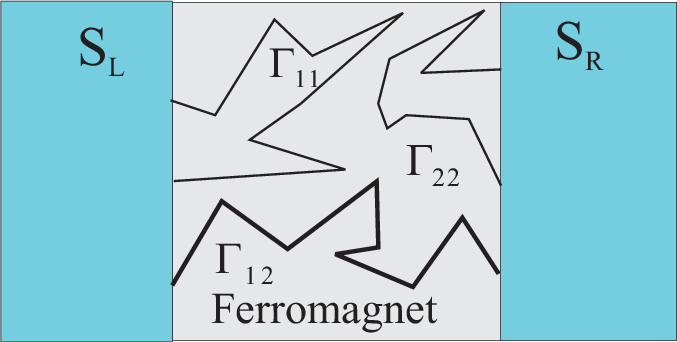}
\caption{(Color online) Random quasiparticle trajectories in the SFS Josephson junction.
} \label{Fig_System}
\end{figure}
The exchange field is responsible for the
 relative phase $\gamma$ gained between the electronic and hole parts of the quasiparticle wave function along these trajectories.
  Averaging the Green functions
 we average in fact the exponential phase factor $e^{i\gamma}$ with the random phase $\gamma$ depending on the trajectory length
 obtaining naturally an exponentially decaying quantity $\propto e^{-x/\xi_f}$, where $x$ is the distance from the SF interface.
  This destructive interference can not play such a dramatic role when we
 calculate root-mean-square (rms) values due to a partial phase gain compensation in squared quantities.
 Considering, e.g.,
 the supercurrent $I$ of the SFS Josephson junction  we can
introduce the rms value of the current as follows:
$\delta I=\sqrt{\langle I^2\rangle-\langle I\rangle^2}$. The compensation of the phase factor $\gamma$ can occur only
for correlated random trajectories passing at the distance not exceeding the Fermi wavelength $\lambda_F=2\pi/k_F$.
This restriction causes the reduction of the $\delta I$ value by a factor of $\sqrt{N}$, where $N$ is the number of transport channels in the junction. Finally, we obtain
$\delta I/\langle I\rangle \sim e^{d/\xi_f}/\sqrt{N}$, where $d$ is the distance between the S electrodes. The number of channels can be of course
pretty large: $N\sim k_F L$ for two dimensional and $N\sim (k_F L)^2$ for three dimensional junctions with the transverse dimension $L$. Nevertheless
the current fluctuations can strongly exceed the average value at large distances $d$ well above the coherence length $\xi_f$.
In this sense these fluctuations are giant compared to the ones in superconductor - normal metal - superconductor (SNS) junctions where the value
$\delta I\sim e\Delta_0/\hbar$ for short junctions with $d\ll\xi_s$ \cite{beenakker} is known to be determined by the universal conductance fluctuations \cite{alt2,lee}
or even smaller for long junctions with $d\gg\xi_s$ \cite{alt}.
Here $\Delta_0$ is the gap in the bulk superconductor and $\xi_s$ is the superconducting coherence length.
Experimentally,
 in each particular sample we can expect to measure a random critical current value which should exhibit giant sample-to-sample fluctuations.
 Thus, in a given experiment one can easily obtain the critical current well above the limit imposed by the Usadel theory which can give us only the average current value. The above arguments and standard Landauer relation between the normal junction resistance $R$ and the $N$ number make it possible to guess a simple estimate for the fluctuating critical current:
 \begin{equation}
 \label{main}
 \delta I \sim \Delta_0/ \sqrt{\hbar R} \ .
 \end{equation}
  Note that this inverse square root dependence differs strongly from the standard relation $I_c\sim \Delta_0/(eR)$ for the SNS junction. Our further calculations nicely confirm the above $\delta I$ estimate and, thus, the observation of this unusual relation between the supercurrent and normal junction resistance could provide a verification of the long range proximity mechanism caused by mesoscopic fluctuations.
  The ensemble averaging laying  in the basis of the derivation of the Usadel equations from the quasiclassical Eilenberger theory overlooks the above fluctuation effects emerging at mesoscopic scales. As we show below these fluctuation effects reveal themselves even in the quasiclassical limit $\lambda_F\rightarrow 0$ when we can neglect the corrections found in  \cite{skvortsov,zyuzin} which vanish in this limit corresponding to a large junction conductance.

We proceed with
a detailed consideration of the critical current fluctuations in the SFS junction using an approach based on the averaging over the random quasiparticle trajectories passing in the field of point scatterers (see \cite{quasiclass} for review). For each random trajectory inside the F layer one can consider the 1D problem for propagating electrons and holes experiencing Andreev
reflection at the point where the trajectory touches the left or right S electrode.
We start from the case $d\ll\xi_s$
and assume the superconducting gap (exchange field)
to vanish inside (outside) the F layer. Thus, we
neglect the so-called inverse proximity effect, i.e., the mutual influence of the order parameters at the interface.
The current -- phase
relation for the short junction limit  can be defined only from the
spectra of the subgap Andreev states at the trajectories ending at both the left and right S electrodes
$\epsilon = \pm\Delta_0 \cos\left( (\varphi \pm \gamma)/2 \right)$ neglecting the contributions from the states above the gap.
Here $\varphi$ is the phase difference between the S electrodes,
and $\pm\gamma$ is the spin-dependent
phase shift between the electron- and hole- like parts of the
total wave function along the quasiclassical trajectory $\Gamma_{12}$ (see Fig.~1).
Each trajectory $\Gamma$ can touch each of the S electrodes only once otherwise part of the trajectory $\Gamma$ touching
the same electrode two times can be considered separately and the corresponding spectrum does not depend on the phase difference $\varphi$
(trajectories $\Gamma_{11}$ and $\Gamma_{22}$ in Fig.~1).
Certainly, there exist trajectories of the length exceeding $\xi_s$ with the quasiparticle spectrum consisting of several subgap branches but
the probability to get such trajectories vanishes for short junctions.
According to the procedure suggested in Ref.~\cite{mel} the phase shift $\gamma$ can be
 determined from the equations which formally coincide with
 the Eilenberger -- type equations written for
 the singlet and triplet parts of the anomalous quasiclassical Green function
$f = f_{sing} + \textbf{f}_t \hat\sigma$ and zero Matsubara frequencies
\begin{equation}
 -i\hbar V_F \partial_s f_{sing}
+2\textbf{h}\textbf{f}_t=0\ ,  -i\hbar V_F\partial_s \textbf{f}_t +2f_{sing}\textbf{h}=0 \,.
\label{S13}
\end{equation}
Here $V_F$ is the Fermi velocity, $s$ is the trajectory coordinate and
the function $f_{sing} (s_R)=\cos\gamma$ taken at the right S
electrode determines the phase gain $\gamma$ along the trajectory.
The boundary conditions at the left electrode read: $f_{sing}(s_L)=1$, $\textbf{f}_t(s_L)=0$.
Let us emphasize here that contrary to the standard consideration
 the Eilenberger-like equations in our approach are written along a random trajectory with many
sharp turns and therefore they do not contain the impurity terms.

Summing up over all trajectories $\Gamma$ we
find the current:
\begin{equation}\label{S2}
 I =\sum\limits_\Gamma
        \left( j(\varphi  + \gamma)
              + j(\varphi-\gamma) \right)(\mathbf{n}_F,\mathbf{n}_L) \ .
\end{equation}
%
%
%
Here $j(\chi)= \sum\limits_{n\ge 1} (j_n/2) \sin (n\chi)$ is the trajectory contribution at zero exchange field,
\begin{equation}\label{S4}
    j_n = \frac{2 e T}{\pi \hbar} \sum_{m=0}^{\infty} \int\limits_0^{2\pi} d\chi
        \frac{\sin\chi\, \sin(n\chi)}{\mu_m + \cos\chi}\,,
\end{equation}
and $\mu_m=2\pi^2 T^2 (2m+1)^2 / \Delta_0^2+1$.
The vectors $\mathbf{n}_L$ and $\mathbf{n}_F$ are the unit vectors
normal to the left electrode surface and parallel to the trajectory direction, respectively.
The vector $\mathbf{n}_F$ parametrizes the trajectories outcoming from the left electrode.
The random phase $\gamma$ depends on the whole path between the electrodes and not just on the distance
between the starting and ending points of the trajectory.
 Taking for simplicity the case of a homogeneous exchange field we find
$\gamma = 2h (s_R-s_L)/\hbar V_F= \Omega t$, where $t$ is the time of flight of electron along the trajectory and $\Omega=2h/\hbar$.

To average the above current expression over the random time of flight $t$
we need to introduce the distribution function describing the probability density
$w({\bf r}_2,{\bf r}_1,t)$
to get the
trajectory starting at a certain point ${\bf r}_1$ at the left electrode at the time $t_1=0$ and touching the right electrode
at an arbitrary point ${\bf r}_2$ at the time $t_2=t$. In the diffusion limit this probability density
is almost independent on the quasiparticle velocity direction at the electrodes and
 satisfies the diffusion equation:
\begin{equation}
\frac{\partial}{\partial t}w=D\frac{\partial^2}{\partial {\bf r}_2^2} w +\delta({\bf r}_2-{\bf r}_1)\delta(t) \ .
\end{equation}
Here we assume the elastic mean free path $\ell$ to be less than all the relevant length scales so that, in particular, $\ell\ll\xi_f$.
The boundary condition should be defined from the fact that the trajectory which touches
the S electrodes should be removed out of game.
An obvious reason is that the corresponding electron moving along the trajectory
 experiences in this case the full Andreev reflection. Thus, at the surfaces of both S electrodes we should put
 $w=0$.
 Choosing ${\bf r}_{1,2}$ at the left and right electrodes, respectively,
 we find
 the probability distribution $P(t)$ for the first-passage time between two electrodes:
  \begin{equation}
 P(t)=-\int\limits_{S_R} D \left({\bf n}_R \frac{\partial}{\partial {\bf r}_R}\right) w({\bf r}_R,{\bf r}_L,t)ds_R \ ,
 \end{equation}
where  the integral is taken over the surface of the right electrode and ${\bf n}_R$ is the unit vector normal to this surface.
The value $P(t)$ gives the probability of the trajectory starting at the point
${\bf r}_L$ at $t_1=0$ to leave the junction in the time interval from $t$ to $t+dt$.
The average current can be written as follows:
\begin{equation}
\langle I\rangle = \sum\limits_{n\ge 1} Nj_n \sin n\varphi
   \langle \cos n\gamma \rangle      \ ,
\end{equation}
where $\langle \cos n\gamma\rangle = Re\int\limits_0^{\infty}e^{-in\Omega t} P(t) dt=Re P(n\Omega)$.
We assume here the surfaces of S electrodes to be
 flat and obtain a one-dimensional problem along the coordinate $x$ perpendicular to these surfaces.
 Introducing the function W(x,t) satisfying the 1D diffusion equation
$D W^{\prime\prime}_{xx}-in\Omega W=0$
with the boundary conditions $DW(x=0)=\ell$ and $W(x=d)=0$ one can find
$P(n\Omega) = DW^\prime_x (x=d,n\Omega)$.
Substituting the solution of the above diffusion equation into the current we obtain:
\begin{equation}
\langle I\rangle= Re \sum\limits_{n\ge 1} Nj_n \sin n\varphi\frac{\ell \sqrt{i n}}{\xi_f}
   \frac{1}{\sinh \left[ \sqrt{i n}d/\xi_f\right]}      \ .
\end{equation}
One can see that this expression reproduces the result of the Usadel theory
only for the first harmonic $I_1\propto \sin\varphi$ in the current - phase relation \cite{buzdin}.
The length $L_n$ of the exponential decay of higher harmonics $I_n\propto \sin n\varphi$ appears to exceed the appropriate length
in the Usadel-type calculation: we obtain here $L_n=\xi_f/\sqrt{n}$ instead of  $L_n=\xi_f/n$.
This result indicates an obvious increase of the range of superconducting correlations due to mesoscopic fluctuations and originates
from the incorrect calculation of the ensemble averages of the product of the anomalous Green functions in the ferromagnet within the Usadel theory.
This failure of the Usadel-type  consideration is caused by the appearance of the random interference phase $\gamma$ and occurs only
in the nonlinear regime of rather strong superconducting correlations.
 Indeed,
considering, e.g., the value   $\langle \cos 2\gamma \rangle$ in the above derivation we calculate the average $\langle |f_{sing}|^2-|f_t|^2 \rangle$
which definitely differs from the product of averages $\langle f_{sing}\rangle\langle f_{sing}^*\rangle-\langle f_t \rangle\langle f_t^* \rangle$.
The mesoscopic fluctuations affect the validity of the nonlinear Usadel equations also for SNS junctions \cite{suppl} providing an exponentially small contribution to the density of states (DOS) below the minigap.
Note that the above approach describes the fluctuation contributions which do not vanish in the limit $\lambda_F \rightarrow 0$ and can, thus,
exceed the sub-minigap DOS corrections found previously in \cite{skvortsov} on the basis of the nonlinear sigma model.
Our contributions are caused by the quantum interference effects associated with a much larger wavelength of the quasiparticle wave function envelope: $\hbar V_F/E$ or $\hbar V_F/h$  for SNS and SFS systems, respectively.


To find the rms value of the supercurrent we evaluate now the expression
\begin{equation}
\langle I^2\rangle=
\sum\limits_{\Gamma,\tilde\Gamma, n,m}
 j_n j_m A_{nm}
(\mathbf{n}_F,\mathbf{n}_L)(\tilde{\bf n}_F,\tilde{\bf n}_L)
\sin n\varphi \sin m\varphi\ ,
\end{equation}
where $A_{nm}=\langle \cos n \Omega t \cos m\Omega \tilde t\rangle$.
The calculation of the above double sum can be done similar to the calculation of the
conductance $R^{-1}=G(d,\ell)$ in a dirty wire above $T_c$.
 Assuming the normal layer thickness to be rather large ($d\gg\xi_f$) and omitting the averages
of the fast oscillating phase factors (which should give the short -- range terms
 decaying at the length $\xi_f$) we get
 \begin{equation}
 \label{rms}
\langle I^2\rangle\simeq (G(\tilde d,\tilde \ell)/4G_0)
\sum\limits_{n\ge 1} j^2_n
 \sin^2 n\varphi\ ,
\end{equation}
 where $G_0=e^2/\pi\hbar$, $\tilde d =\Omega d/k_FV_F$ and $\tilde\ell=\Omega\ell/k_F V_F$.
Taking the Drude-type conductance
$G/G_0 = N\ell/d$
for a disordered wire of the length $d$  we find
the estimate
\begin{equation}
\label{rms1}
\sqrt{\langle I^2\rangle- \langle I\rangle^2}\sim
\sqrt{\frac{N\ell}{ d}} \sqrt{\sum\limits_{n\ge 1} j^2_n
 \sin^2 n\varphi}\ .
\end{equation}

 The deviations from the Drude result arise naturally from the so-called interference or
localization corrections to the conductance \cite{quasiclass}. Perturbatively, they can be estimated as terms arising from the paths with self-crossings
in the above double sum over the trajectories. According to the Thouless criterion \cite{thouless} the localization effects in a disordered wire are small
provided the effective number $N\ell/d$ of the conducting modes is large. Thus, one can expect our Drude-type estimate to hold in the case
$N\ell/d\gg 1$. In the opposite limit the wire conductance in Eq.~(\ref{rms}) and, thus, the rms value of the critical current
decay exponentially at the length $N\ell$.

%
%
%

Comparing the rms value with the average current taken in the same limit $d\gg\xi_f$
we find
\begin{equation}
\label{fluct}
\delta I/\langle I\rangle\sim
\sqrt{\frac{\xi_f^2}{N\ell d}} \exp \left(\frac{d}{\xi_f\sqrt{2}
}\right)\ .
\end{equation}
This expression for current fluctuations definitely can not be obtained within the averaged Usadel theory and  results from the partial
 cancelation
of the interference contributions in the product of the anomalous Green functions.
Note that turning to the limit $d\ll\xi_f$, i.e., to the case
of the SNS junction
our consideration should give a vanishing $\delta I$ value since we disregarded the quantum interference of random semiclassical trajectories
responsible for standard mesoscopic fluctuations \cite{beenakker}.
 Despite the small factor $N^{-1/2}$ in Eq.~\ref{fluct}
the current fluctuations for $d\gg\xi_f$ appear to be giant compared to the current average value which decays exponentially at the small distance $\xi_f$.
The rms  value  can well exceed the Josephson current quantum $e\Delta/\hbar$ in
SNS
junctions \cite{beenakker}. It is also important to note that contrary to the average current the fluctuating contributions to higher harmonics of the current -- phase relation are not suppressed exponentially compared to the first harmonic. This strong anharmonicity probably relates to the experimental data on the  large  second harmonics in SFS junctions \cite{ryazanov,blamire}. Certainly, in realistic junctions the above assumption of the full Andreev reflection at the SF boundaries can be broken due to the effect of the interface potential barriers
which certainly suppress the higher current harmonics.
 Still the main effect, namely, the partial compensation of the phases $\gamma$ in the rms values should exist even in the presence of the barriers.


 One can easily see that the above long range behavior of the critical current fluctuations holds also
 beyond the short junction approximation (i.e. for $d>\xi_s$)
at least for the first harmonic $I_1(\varphi)$. Indeed, the critical current in this limit is determined by the
singlet component of the anomalous Green function  $\sum\limits_\Gamma f_{sing\Gamma}=\sum\limits_\Gamma \cos\gamma_\Gamma$. The average current, therefore, decays exponentially as $\langle I\rangle \propto (\ell N/\xi_f) e^{-d/\xi_f\sqrt{2}}$ while the rms average becomes long ranged because of the partial phase compensation at close trajectories:  $\langle(\delta I)^2 \rangle\propto\langle f_{sing}^2\rangle \propto N\ell/d$.
Thus, the above calculations confirm the estimate (\ref{main}) both for short and long junctions. Certainly, further increase in the distance $d$ will give us the exponential decay of the supercurrent but at the distances $\gtrsim\xi_N$.

The mesoscopic fluctuations considered in our work should be most easily observed in
ferromagnetic wires because their relative contribution decays with the increase of the $N$ number.
Let us perform the quantitative comparison of our theory with the
experimental data  of the Ref.~\cite{wang} which has reported on
the observation of the critical current $I_{c}\sim 10\mu A$
for the Co  nanowire with the length $d=0.6\mu m$
and diameter $2r=40nm$ at temperature $T=1.8$ $K$.
This temperature is much smaller than the critical
temperature of W electrodes $T_{c}\sim 5$ $K$ so we may use the low
temperature limit for $j_{n}$ and the Eq.~(\ref{rms}) transforms into the estimate:
$\delta I\sim\Delta _{0}/2\sqrt{\pi\hbar R}$. We
take the resistivity  $\rho =10$ $\mu \Omega $ $cm$ \ obtained in [13] for the
Co nanowire with the same diameter and the length $d=1.5\mu m$ and
find $R\sim 4\Omega $.
Taking the gap $\Delta _{0}\sim 1.74T_{c}\sim 8\ K$
 we finally get the value $\delta I\sim 1 \mu A$ which is only an order of magnitude less that the critical current observed in Ref. \cite{wang}.
The remaining discrepancy is probably caused by the overestimating of the wire resistance $R$ due to the presence of contact resistances due to the
 interdiffusion of W at the distance $\sim 200nm$.
It is also useful to compare the fluctuation contribution with possible effect of spin-triplet correlations which can still appear, e.g.,
due to some noncollinearity of magnetic moments at the interfaces. Introducing such thin noncollinear domains of the thickness $d_i\ll\xi_f$
at the left and right ends of the ferromagnetic wire one obtains the long-range current contribution in the form \cite{houset-buzdin}:
 $I_{tr}\sim (d_i/\xi_f)^4 \Delta_0/e R$. One can see that for the wires with rather large resistances $R>G_0^{-1} (d_i/\xi_f)^4$ the fluctuation
 contribution dominates.

%

We now briefly comment on the effect of mesoscopic fluctuation on the local DOS (LDOS) at the Fermi level.
In the ballistic system for straight linear trajectories one can easily obtain an Eilenberger-type
 expression for this quantity as a sum of contributions from different quasiclassical paths. This expression can be simplified applying
 the normalization condition for quasiclassical Green functions and taking the perturbation expansion in powers of the $f$
function (see, e.g., \cite{eschrig2} for convenient notations).
Generalizing this expressions for the trajectories experiencing many sharp turns one can get:
  $\delta\nu /\nu_F \propto -N^{-1}\sum\limits_\Gamma (|f_{sing}|^2-|f_t|^2)$, where $\nu_F$ is the normal metal LDOS.
 The ensemble average of this value certainly decays exponentially $\langle\delta\nu /\nu_F \rangle\propto-\langle\cos 2\gamma \rangle\propto - (\ell/\xi_f)e^{-d/\xi_f}\cos(d/\xi_f+\pi/4)$ with the increase in the distance $d$ from the S electrode.
 The fluctuating LDOS contains a long range contribution similar to the one calculated above for the critical current:
 $\sqrt{\langle(\delta\nu /\nu_F)^2 \rangle} \propto \sqrt{\ell/dN}$. This nonexponential behavior of the fluctuating superconducting contribution to the LDOS could be measured by a local conductance probe at different points of a ferromagnetic nanowire placed in contact to a superconductor
  providing, thus, a possible explanation of
 the long range proximity effect observed in Refs.~\cite{giroud,petrashov,nugent}.

The direct observation of the giant sample-to-sample fluctuations assumes the measurements of the critical current or LDOS on different junctions. It would be much more convenient to find the way
 to change the interference phases $\gamma$ in a given sample and measure the junction `` fingerprints '' in analogy to the observation of universal conductance fluctuations vs applied magnetic field \cite{akker}. Indeed, such type of experiment in the SFS junctions may become possible provided we apply the magnetic field which can affect the domain structure in the F layer without producing non-collinear magnetic regions  to avoid the admixture of the long-range triplet correlations.

To sum up, we suggest a  theoretical model describing the mesoscopic fluctuations in SF and SN systems. This model allows to obtain a
huge fluctuation contribution in supercurrent in SFS junctions and LDOS which survives the destructive effect of the exchange field and could, thus, provide an explanation
of the experimental data on the long-range proximity phenomena in ferromagnetic wires even in the absence of the exchange field inhomogeneity.
 The resulting fluctuating supercurrent is a sign-changing quantity and, thus, the ensemble fluctuations cause the appearance of both zero and $\pi$- junctions.
 Our analysis reveals also that fluctuations can be responsible for anomalously large values of the second and higher harmonics in the current-phase relation.

This work
was supported by the French ANR "MASH," NanoSC
COST Action MP1201,
and Russian
Science Foundation under Grant No. 15-12-10020 (ASM).

\widetext

\appendix

\section{Supplementary material for ``Giant Mesoscopic Fluctuations and Long Range Superconducting Correlations
 in Superconductor--Ferromagnet structures''. Density of states in SNS junctions}
To illustrate the mechanism of the random phase accumulation and extend our discussion of the Usadel theory validity
we have compared its results \cite{golubov,bruder1,zhou,bruder2,altland}
with the ones found by the above method of averaging over the random trajectories for the dependence of density of states (DOS)
on energy $E$ in SNS junctions. In this auxiliary problem the random phase gain occurs at the trajectory length $\sim\hbar V_F/E$. The behavior of the local DOS
in the middle of the junction obtained by different method appears to be in a good agreement for  energies $E$  exceeding the Thouless energy
$E_{Th}=\hbar D/d^2$, i.e. for a small anomalous Green function described by the linearized Usadel theory.
Below the minigap $E<3.12 E_{Th}$ our approach gives a finite fluctuation DOS contribution which vanishes exponentially
for $E\rightarrow 0$. Note that in the limit
$\lambda_F \rightarrow 0$ the fluctuation corrections found previously in \cite{skvortsov} on the basis of the nonlinear sigma model should vanish.
Nevertheless, in our approach we still get the fluctuation corrections caused by
 the quantum interference effects associated with a much larger wavelength of the quasiparticle wave function envelope: $\hbar V_F/E$ or $\hbar V_F/h$
 for SNS and SFS systems, respectively.
  These large scale interference effects reveal themselves in the regime of strong superconducting correlations, i.e. below the minigap in the SNS junctions and for higher harmonics in Josephson relation for SFS junctions.

Here we apply the approach based on the averaging over the quasiclassical trajectories for the calculation
of the local density of states (LDOS) in the middle plane $x=0$ of the superconductor -- normal metal -- superconductor junction (see Fig.~1).
\begin{figure}[b!]
\includegraphics[width=0.3\textwidth]{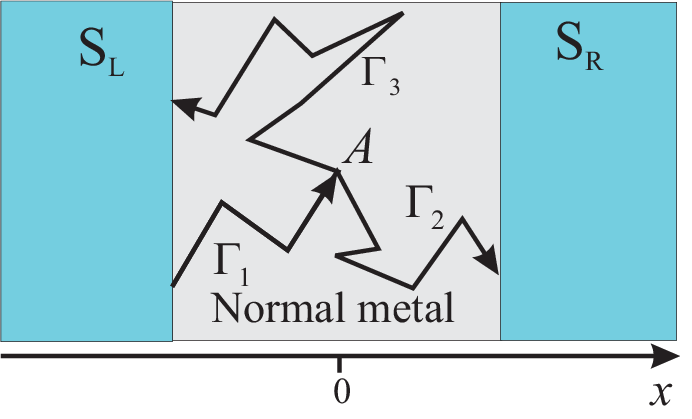}
\caption{(Color online) Random quasiparticle trajectories in the SNS system.
}
\end{figure}
Let us consider a trajectory passing through  a certain point $A$ of this plane and consisting of two parts which start at the point $A$ and
end either at the left or right superconducting lead. The low energy subgap spectrum for quasiparticles moving along this trajectory is quantized due to the Andreev reflection at the trajectory ends: $\varepsilon = \pi (n+1/2)\hbar V_F/L$, where $L$ is the trajectory length, $V_F$ is the Fermi velocity and $n$ is an integer. The length $L$ can be written in the form $L=V_F (t_1+t_2)$ where the times $t_1$ and $t_2$ are needed for a
quasiparticle diffusing along different paths to get the superconducting electrodes from the point $A$. Taking account of the wave function normalization
and assuming the times $t_{1,2}$ to be statistically independent one can get the local density of states averaged over the random paths
and normalized to the local density of states in the normal metal:
\begin{equation}
N(E)= \sum\limits_{n}
\int\limits_{0}^{\infty} dt_1
\int\limits_{0}^{\infty} dt_2
P(t_1)P(t_2)  \frac{\pi\hbar}{(t_1+t_2)}\delta\left(E - \frac{\pi \hbar(n+1/2)}{t_1+t_2}\right) \ ,
\end{equation}
where the distribution function $P(t)$ is defined in the main text.
The Fourier transform of the function $P(t)$  reads
\begin{equation}
P(\omega)=\frac{1}{\cos\sqrt{-i\omega/4E_{Th}}} \ ,
\end{equation}
where $E_{Th}=\hbar D/d^2$ is the Thouless energy and $d$ is the distance between the superconducting electrodes.
Evaluating the above integral we obtain:
\begin{equation}
N(E)=\frac{4\pi E_{Th}}{E} \sum\limits_{n=0}^{\infty} \frac{\frac{2a_n E_{Th}}{E}\cosh\frac{a_n E_{Th}}{E}-
\sinh\frac{a_nE_{Th}}{E}}
{\sinh^2\frac{a_nE_{Th}}{E}} \ ,
\end{equation}
where $a_n=\pi^3 (2n+1)^2/2$.
For the energies well above the Thouless energy the density of states approaches unity:
\begin{equation}
N(E)\rightarrow \sum\limits_{n=0}^{\infty} \frac{8}
{\pi^2 (2n+1)^2}=1 \ ,
\end{equation}
For low energies $E\ll E_{Th}$ we find:
\begin{equation}
N(E)\simeq\frac{16\pi E^2_{Th}}{E^2} \sum\limits_{n=0}^{\infty}
a_n e^{-a_nE_{Th}/E}
\ .
\end{equation}
Neglecting this exponentially small contribution we restore the standard result of the Usadel theory, namely,
zero density of states below the minigap $E<3.12 E_{Th}$ \cite{golubov,bruder1,zhou,bruder2,altland}.
 The overall behavior of the energy dependence of the
density of states is shown in Fig.~2. This behavior is rather close to the one resulting from the Usadel equations \cite{golubov,bruder1,zhou,bruder2,altland}.
Indeed, one can see that our calculations reproduce, in particular, the peak in the local DOS and
the peak amplitude is close to the one found in Usadel-type calculations.

\begin{figure}[h!]
\includegraphics[width=0.5\textwidth]{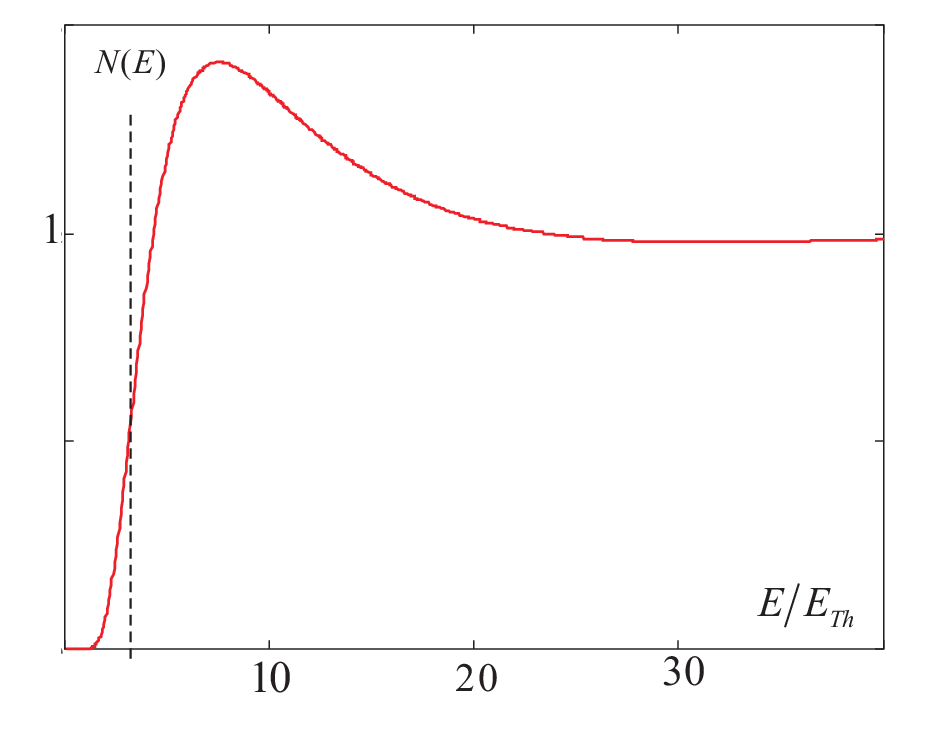}
\caption{(Color online) Local density of states in the middle point A of the SNS junction.
The vertical dashed line corresponds to the value of the spectral gap $E/E_{Th}=3.12$ in the Usadel theory.}
\end{figure}

%
%
%
%
%
%
%
%

%
\end{document}